\documentstyle[12pt,twoside]{article}

\catcode`@=11
\def\citer{\@ifnextchar [{\@tempswatrue\@citexr}{\@tempswafalse\@citexr[]}}
 
\def\@citexr[#1]#2{\if@filesw\immediate\write\@auxout{\string\citation{#2}}\fi
  \def\@citea{}\@cite{\@for\@citeb:=#2\do
    {\@citea\def\@citea{--\penalty\@m}\@ifundefined
       {b@\@citeb}{{\bf ?}\@warning
       {Citation `\@citeb' on page \thepage \space undefined}}%
\hbox{\csname b@\@citeb\endcsname}}}{#1}}
\catcode`@=12

\def\refeq#1{\mbox{eq.~(\ref{#1})}}

\def\citere#1{\mbox{Ref.~\cite{#1}}}
\def\citeres#1{\mbox{Refs.~\cite{#1}}}

\newcommand{\mst}{m_{\tilde{t}}}
\newcommand{\delmst}{\Delta\mst}

\newcommand{\mste}{m_{\tilde{t}_1}}
\newcommand{\mstz}{m_{\tilde{t}_2}}

\newcommand{\MstL}{M_{\tilde{t}_L}}
\newcommand{\MstR}{M_{\tilde{t}_R}}
\newcommand{\MsbL}{M_{\tilde{b}_L}}
\newcommand{\MsbR}{M_{\tilde{b}_R}}

\newcommand{\Mtlr}{M_{t}^{LR}}

\newcommand{\Pe}{\phi_1}
\newcommand{\Pz}{\phi_2}

\newcommand{\PePz}{\phi_1\phi_2}
\newcommand{\mpe}{m_{\Pe}}
\newcommand{\mpz}{m_{\Pz}}
\newcommand{\mpez}{m_{\PePz}}

\newcommand{\SU}{\mathrm {SUSY}}

\newcommand{\oaas}{{\cal O}(\alpha\alpha_s)}
\newcommand{\cp}{{\cal CP}}

\newcommand{\edz}{\frac{1}{2}}
\newcommand{\twol}{two-loop}
\newcommand{\onel}{one-loop}
\newcommand{\mma}{{\em Mathematica}}

\newcommand{\fh}{{\em FeynHiggs}}

\newcommand{\MZ}{M_Z}
\newcommand{\MA}{M_A}
\newcommand{\mh}{m_h}
\newcommand{\mH}{m_H}

\newcommand{\dr}{\De\rho}
\newcommand{\mt}{m_{t}}
\newcommand{\mtms}{\overline{m}_t}

\newcommand{\mb}{m_{b}}

\newcommand{\mgl}{m_{\tilde{g}}}

\newcommand{\Stop}{\tilde{t}}
\newcommand{\StopL}{\tilde{t}_L}
\newcommand{\StopR}{\tilde{t}_R}
\newcommand{\Stope}{\tilde{t}_1}
\newcommand{\Stopz}{\tilde{t}_2}

\newcommand{\Sbot}{\tilde{b}}

\newcommand{\tst}{\theta_{\tilde{t}}}
\newcommand{\tsb}{\theta_{\tilde{b}}}
\newcommand{\tsf}{\theta\kern-.20em_{\tilde{f}}}
\newcommand{\tsfp}{\theta\kern-.20em_{\tilde{f}\prime}}
\newcommand{\tsq}{\theta\kern-.15em_{\tilde{q}}}
\newcommand{\sw}{s_W}
\newcommand{\cw}{c_W}

\newcommand{\sintt}{\sin\tst}

\newcommand{\sinQZtt}{\sin^2 2\tst}
\newcommand{\sintb}{\sin\tsb}

\newcommand{\costt}{\cos\tst}

\newcommand{\KL}{\left(}
\newcommand{\KR}{\right)}
\newcommand{\KKL}{\left[}
\newcommand{\KKR}{\right]}

\newcommand{\VL}{\left( \begin{array}{c}}
\newcommand{\VR}{\end{array} \right)}
\newcommand{\ML}{\left( \begin{array}{cc}}
\newcommand{\MLd}{\left( \begin{array}{ccc}}
\newcommand{\MLv}{\left( \begin{array}{cccc}}
\newcommand{\MR}{\end{array} \right)}

\newcommand{\hc}{\mbox {h.c.}}

\newcommand{\Tb}{\tan \beta\hspace{1mm}}

\newcommand{\CTb}{\cot \beta\hspace{1mm}}

\newcommand{\Sb}{\sin \beta\hspace{1mm}}
\newcommand{\SQb}{\sin^2\beta\hspace{1mm}}

\newcommand{\Cb}{\cos \beta\hspace{1mm}}
\newcommand{\CQb}{\cos^2\beta\hspace{1mm}}

\newcommand{\Sa}{\sin \alpha\hspace{1mm}}
\newcommand{\SQa}{\sin^2\alpha\hspace{1mm}}
\newcommand{\Ca}{\cos \alpha\hspace{1mm}}
\newcommand{\CQa}{\cos^2\alpha\hspace{1mm}}

\newcommand{\CZb}{\cos 2\beta\hspace{1mm}}

\newcommand{\gev}{\,\, {\mathrm GeV}}

\newcommand{\BC}{\begin{center}}
\newcommand{\EC}{\end{center}}
\newcommand{\BE}{\begin{equation}}
\newcommand{\EE}{\end{equation}}
\newcommand{\BEA}{\begin{eqnarray}}
\newcommand{\BEAnn}{\begin{eqnarray*}}
\newcommand{\EEA}{\end{eqnarray}}
\newcommand{\EEAnn}{\end{eqnarray*}}
\newcommand{\non}{\nonumber}
\newcommand{\id}{{\rm 1\kern-.12em
\rule{0.3pt}{1.5ex}\raisebox{0.0ex}{\rule{0.1em}{0.3pt}}}}

\def\al{\alpha}
\def\als{\alpha_s}

\def\De{\Delta}

\def\hSi{\hat{\Sigma}}

\def\hSii{\hat{\Sigma}^{(i)}}
\def\hSie{\hat{\Sigma}^{(1)}}
\def\hSiz{\hat{\Sigma}^{(2)}}

\def\draftdate{\relax}
\def\mda{\relax}
\def\mua{\relax}
\def\mla{\relax}
\def\draft{
\def\thtystars{******************************}
\def\sixtystars{\thtystars\thtystars}
\typeout{}
\typeout{\sixtystars**}
\typeout{* Draft mode!
         For final version remove \protect\draft\space in source file *}
\typeout{\sixtystars**}
\typeout{}
\def\draftdate{\today}
\def\mua{\marginpar[\boldmath\hfil$\uparrow$]%
                   {\boldmath$\uparrow$\hfil}%
                    \typeout{marginpar: $\uparrow$}\ignorespaces}
\def\mda{\marginpar[\boldmath\hfil$\downarrow$]%
                   {\boldmath$\downarrow$\hfil}%
                    \typeout{marginpar: $\downarrow$}\ignorespaces}
\def\mla{\marginpar[\boldmath\hfil$\rightarrow$]%
                   {\boldmath$\leftarrow $\hfil}%
                    \typeout{marginpar: $\leftrightarrow$}\ignorespaces}
\def\Mua{\marginpar[\boldmath\hfil$\Uparrow$]%
                   {\boldmath$\Uparrow$\hfil}%
                    \typeout{marginpar: $\Uparrow$}\ignorespaces}
\def\Mda{\marginpar[\boldmath\hfil$\Downarrow$]%
                   {\boldmath$\Downarrow$\hfil}%
                    \typeout{marginpar: $\Downarrow$}\ignorespaces}
\def\Mla{\marginpar[\boldmath\hfil$\Rightarrow$]%
                   {\boldmath$\Leftarrow $\hfil}%
                    \typeout{marginpar: $\Leftrightarrow$}\ignorespaces}
\overfullrule 5pt
\oddsidemargin -15mm
\marginparwidth 29mm
}

\begin{document}
\thispagestyle{empty}

\def\thefootnote{\fnsymbol{footnote}}

\begin{flushright}
KA-TP-16-1998\\
DESY 98-193\\
CERN-TH/98-389\\
hep-ph/9812320 \\
\end{flushright}

\vspace{1cm}

\begin{center}

{\large\sc {\bf {\em FeynHiggs}: a program for the calculation }}

\vspace*{0.4cm} 

{\large\sc {\bf of the masses of the neutral $\cp$-even }}

\vspace*{0.4cm} 

{\large\sc {\bf Higgs bosons in the MSSM }}

\vspace{1cm}

{\sc 
S.~Heinemeyer$^{1}$%
\footnote{email: Sven.Heinemeyer@desy.de}%
, W.~Hollik$^{2,3}$%
\footnote{email: Wolfgang.Hollik@physik.uni-karlsruhe.de}%
, G.~Weiglein$^{3}$%
\footnote{email: georg@particle.physik.uni-karlsruhe.de}
}

\vspace*{1cm}

{\sl
$^1$ DESY Theorie, Notkestr. 85, 22603 Hamburg, Germany

\vspace*{0.4cm}

$^2$ Theoretical Physics Division, CERN, CH-1211 Geneva 23, Switzerland

\vspace*{0.4cm}

$^3$ Institut f\"ur Theoretische Physik, Universit\"at Karlsruhe, \\
D--76128 Karlsruhe, Germany
}

\end{center}

\vspace*{1cm}

\begin{abstract}
\fh\ is a Fortran code for the calculation of the masses of the
neutral $\cp$-even Higgs bosons in the MSSM up to \twol\ order. It is based
on the complete diagrammatic on-shell results at the \onel\ level,
the leading diagrammatic \twol\ QCD contributions and further improvements
taking into account leading electroweak \twol\ and
leading higher-order QCD corrections. 
The Higgs-boson masses are calculated as functions of the MSSM
parameters for general mixing in the
scalar top sector and 
arbitrary choices of the parameters in the Higgs sector of the model.
\end{abstract}

\def\thefootnote{\arabic{footnote}}
\setcounter{page}{0}
\setcounter{footnote}{0}

\newpage


\section{Introduction}

A direct and very 
stringent test of Supersymmetry (SUSY) is provided by
the search for the lightest Higgs boson, since the prediction of a
relatively 
light Higgs boson is common to all supersymmetric models whose
couplings remain in the perturbative regime up to a very high energy
scale~\cite{susylighthiggs}.
A precise prediction for the mass of the lightest Higgs boson, $\mh$,
in terms 
of the relevant SUSY parameters is crucial in order to determine the
discovery and exclusion potential of LEP2 and the upgraded Tevatron.
If the Higgs boson exists, it will be accessible at the LHC and future
linear colliders, where then a high-precision
measurement of the mass of this particle will become feasible.
A precise knowledge of the mass of the heavier $\cp$-even Higgs boson,
$\mH$, is important for
resolving the mass splitting between the $\cp$-even and -odd
Higgs-boson masses.

In the Minimal Supersymmetric Standard Model (MSSM)~\cite{mssm} the
mass of the lightest Higgs boson is restricted at the tree
level to be smaller
than the $Z$-boson mass. This bound, however, is strongly affected by
the inclusion of radiative corrections, 
which yield an upper bound of about $130 \gev$%
~\cite{mhiggs1l,mhiggs1lfullb,mhiggs1lfull,mhiggs2la,mhiggs2lb,mhiggsRG,mhiggsEffPot}. 
Results beyond \onel\ order have been obtained in several
approaches: 
a Feynman diagrammatic calculation of the leading QCD
corrections has been performed~\cite{mhiggs2la,mhiggs2lb}.
Also renormalization group (RG)
methods have been applied in order to obtain leading logarithmic
higher-order contributions~\cite{mhiggsRG}.
Furthermore the leading \twol\ QCD corrections have been calculated in the
effective potential method~\cite{mhiggsEffPot}.
All these calculations show that the corrections beyond \onel\ order
lead to a sizable decrease of $\mh$
of up to $20 \gev$.

Concerning the calculation of the lighter and heavier neutral
$\cp$-even Higgs bosons two different
kinds of computer codes have been used for phenomenological analyses
so far: they are either based on the RG improved \onel\ effective 
potential approach~\cite{mhiggsRG} or on the \onel\ diagrammatic
on-shell calculation~\cite{mhiggs1lfullb,mhiggs1lfull}. These
approaches differ by ${\cal O}(10 \gev)$.

\smallskip
Here we present a new Fortran program named \fh, which is based on
the results of the Feynman-diagrammatic calculations up to $\oaas$ given in
Refs.~\cite{mhiggs1lfull,mhiggs2la,mhiggs2lb}. It includes the
complete diagrammatic on-shell results at the \onel\ level,
the leading diagrammatic \twol\ QCD contributions and further improvements
taking into account leading electroweak \twol\ and
leading higher-order QCD corrections. 
The calculation of $\mh$ and $\mH$ is performed for arbitrary
values of parameters in the $t-\Stop$-sector and
the Higgs sector of the MSSM. As a subroutine, \fh\ can be linked to other 
programs thus incorporating in an easy way a precise prediction for
$\mh$ and $\mH$.
In addition the program provides as an option the calculation of the SUSY
contribution to the $\rho$-parameter in $\oaas$, based
on~\citeres{drhosuqcd,precobssusyqcd}. In this way 
experimentally disfavored combinations of squark masses can
automatically be excluded.

\smallskip
The paper is organized as follows: in section~2 we specify our
notations and give a brief outline of the calculation of $\mh$ and $\mH$. In
section~3 the program \fh\ is described in detail. Examples
of how to use \fh\ are shown in section~4. The conclusions are given
in section~5.


\section{The calculational basis}

\subsection{The top-squark sector of the MSSM}

In order to fix the notation we shortly list our conventions for the
MSSM scalar top sector:
the mass matrix in the basis of the current eigenstates $\StopL$ and
$\StopR$ is given by
\BE
\label{stopmassmatrix}
{\cal M}^2_{\Stop} =
  \ML \MstL^2 + \mt^2 + \CZb (\edz - \frac{2}{3} \sw^2) \MZ^2 &
      \mt \Mtlr \\
      \mt \Mtlr &
      \MstR^2 + \mt^2 + \frac{2}{3} \CZb \sw^2 \MZ^2 
  \MR,
\EE
where 
\BE
\mt \Mtlr = \mt (A_t - \mu \CTb)~.
\EE
Diagonalizing the $\Stop$-mass matrix yields the mass eigenvalues 
$\mste, \mstz$ and the $\Stop$~mixing angle~$\tst$, which relates
the current eigenstates to the mass eigenstates:
\BE
\VL \Stope \\ \Stopz \VR = \ML \costt & \sintt \\ -\sintt & \costt \MR 
                       \VL \StopL \\ \StopR \VR~.
\label{sfermrotation}
\EE


\subsection{Calculation of the Higgs-boson masses}
\label{subsec:mhcalc}

Contrary to the Standard Model (SM), in the MSSM two Higgs doublets
are required.
The  Higgs potential~\cite{hhg}
\BEA
\label{Higgspot}
V &=& m_1^2 H_1\bar{H}_1 + m_2^2 H_2\bar{H}_2 - m_{12}^2 (\epsilon_{ab}
      H_1^aH_2^b + \hc)  \nonumber \\
   && \mbox{} + \frac{g'^2 + g^2}{8}\, (H_1\bar{H}_1 - H_2\bar{H}_2)^2
      +\frac{g^2}{2}\, |H_1\bar{H}_2|^2
\EEA
contains $m_1, m_2, m_{12}$ as soft SUSY breaking parameters;
$g, g'$ are the $SU(2)$ and $U(1)$ gauge couplings, and 
$\epsilon_{12} = -1$.

The doublet fields $H_1$ and $H_2$ are decomposed  in the following way:
\BEA
H_1 &=& \VL H_1^1 \\ H_1^2 \VR = \VL v_1 + (\phi_1^{0} + i\chi_1^{0})
                                 /\sqrt2 \\ \phi_1^- \VR ,\non \\
H_2 &=& \VL H_2^1 \\ H_2^2 \VR =  \VL \phi_2^+ \\ v_2 + (\phi_2^0 
                                     + i\chi_2^0)/\sqrt2 \VR.
\label{eq:hidoubl}
\EEA
The potential (\ref{Higgspot}) can be described with the help of two  
independent parameters (besides $g$ and $g'$): 
$\Tb = v_2/v_1$ and $M_A^2 = -m_{12}^2(\Tb+\CTb)$,
where $M_A$ is the mass of the $\cp$-odd $A$ boson.

In order to obtain the $\cp$-even neutral mass eigenstates, the rotation 
\BEA
\VL H^0 \\ h^0 \VR &=& \ML \Ca & \Sa \\ -\Sa & \Ca \MR 
\VL \phi_1^0 \\ \phi_2^0 \VR  
\label{higgsrotation}
\EEA
is performed, where the mixing angle $\alpha$ is given in terms of
$\Tb$ and $M_A$ as follows:
\BE
\tan 2\alpha = \tan 2\beta \frac{\MA^2 + \MZ^2}{\MA^2 - \MZ^2},
\quad - \frac{\pi}{2} < \alpha < 0. 
\EE


\bigskip
At tree level the mass matrix of the neutral $\cp$-even Higgs bosons
in the $\phi_1-\phi_2$ basis can be expressed 
in terms of $\MZ$ and $\MA$ as follows:
\BEA
M_{\rm Higgs}^{2, {\rm tree}} &=& \ML \mpe^2 & \mpez^2 \\ 
                           \mpez^2 & \mpz^2 \MR \non\\
&=& \ML \MA^2 \SQb + \MZ^2 \CQb & -(\MA^2 + \MZ^2) \Sb \Cb \\
    -(\MA^2 + \MZ^2) \Sb \Cb & \MA^2 \CQb + \MZ^2 \SQb \MR,
\EEA
which by diagonalization 
according to \refeq{higgsrotation} yields the tree-level
Higgs-boson masses
\BE
M_{\rm Higgs}^{2, {\rm tree}} 
   \stackrel{\al}{\longrightarrow}
   \ML m_{H,{\rm tree}}^2 & 0 \\ 0 &  m_{h,{\rm tree}}^2 \MR.
\EE

\bigskip
In the Feynman-diagrammatic approach the 
Higgs-boson masses in higher orders are derived by finding the poles
of the $h-H$-propagator matrix whose inverse reads:
\BE
\left(\Delta_{\rm Higgs}\right)^{-1}
= - i \ML q^2 -  m_{H,{\rm tree}}^2 + \hSi_{H}(q^2) &  \hSi_{hH}(q^2) \\
     \hSi_{hH}(q^2) & q^2 -  m_{h,{\rm tree}}^2 + \hSi_{h}(q^2) \MR~.
\label{higgsmassmatrixnondiag}
\EE
The poles are then obtained by solving the equation
\BE
(q^2 - m_{h,{\rm tree}}^2 + \hSi_{h}(q^2))
(q^2 - m_{H,{\rm tree}}^2 + \hSi_{H}(q^2))
-(\hSi_{hH}(q^2))^2 = 0~.
\label{higgsmasseq}
\EE

In the following the $\hSii$ denote the \onel\ ($i=1$) and the \twol\
($i=2$) contributions to the renormalized self-energies.

In \fh\ the \onel\ results for the Higgs-boson self-energies $\hSie_s(q^2)$
are calculated according 
to \citere{mhiggs1lfull}. They contain the full \onel\ contribution
obtained via an explicit Feynman-diagrammatic 
calculation in the on-shell scheme. Here the gaugino parameters $M_1$
and $M (\equiv M_2)$ enter in 
the neutralino mass matrix. $M_1$ is fixed via the GUT relation
\BE
M_1 = \frac{5}{3} \frac{\sw^2}{\cw^2} M,
\EE
whereas $M$ is kept as a free input parameter.

The \twol\ results for the Higgs-boson self-energies $\hSiz_s$ are taken from
Refs.~\cite{mhiggs2la,mhiggs2lb}. The leading \twol\ corrections
have been obtained by calculating the $\oaas$ contribution of the
$t-\Stop$-sector to the renormalized Higgs-boson self-energies at zero
external momentum from the Yukawa part of the theory. These 
\twol\ QCD corrections are
expected to constitute the most sizable part of the full set of \twol\
corrections.
In Refs.~\cite{mhiggs2la,mhiggs2lb} the self-energies have
been computed first in the 
$\Pe-\Pz$ basis and afterwards rotated into the $h-H$ basis
according to \refeq{higgsrotation}:
\BEA
\hSiz_{H} &=& \CQa \hSiz_{\Pe} + \SQa \hSiz_{\Pz} + 
              2 \Sa \Ca \hSiz_{\PePz} \non \\
\hSiz_{h} &=& \SQa \hSiz_{\Pe} + \CQa \hSiz_{\Pz} - 
              2 \Sa \Ca \hSiz_{\PePz} \non \\
\hSiz_{hH} &=& - \Sa \Ca \KL \hSiz_{\Pe} - \hSiz_{\Pz} \KR + 
              (\CQa - \SQa) \hSiz_{\PePz} .
\label{higgsserotation}
\EEA

Thus for our results up to the \twol\ level the
matrix~(\ref{higgsmassmatrixnondiag})  
contains the renormalized Higgs-boson self-energies
\BE
\hSi_s(q^2) = \hSie_s(q^2) + \hSiz_s(0), \quad s = h, H, hH,
\label{renhiggsse}
\EE
where the momentum dependence is neglected only in the \twol\
contribution.
The calculation is performed 
for arbitrary parameters of the Higgs and the scalar top sector and
for arbitrary gluino mass~$\mgl$.
Thus the accuracy of the calculation does not
depend on how the parameters of the $\Stop$ sector 
$\mste, \mstz$ and $\tst$ are chosen.

\bigskip
In order to take into account the leading electroweak \twol\
contribution to the mass of the lightest Higgs boson,
we have implemented the leading Yukawa correction of 
${\cal O}(G_F^2 \mt^6)$, which gives a sizable contribution only for $\mh$.
The formula is taken over from the result obtained by renormalization
group methods. It reads~\cite{mhiggsRGmhref}
\BEA
\label{yukawaterm}
\Delta\mh^2 &=& \frac{9}{16\pi^4} G_F^2 \mt^6
               \KKL \tilde{X} t + t^2 \KKR \\
\mbox{with} && \tilde{X} = \Bigg[
                \KL \frac{\mstz^2 - \mste^2}{4 \mt^2} \sinQZtt \KR^2
                \KL 2 - \frac{\mstz^2 + \mste^2}{\mstz^2 - \mste^2}
                      \log\KL \frac{\mstz^2}{\mste^2} \KR \KR \non\\
            && \mbox{}\hspace{1cm}  
               + \frac{\mstz^2 - \mste^2}{2 \mt^2} \sinQZtt
                      \log\KL \frac{\mstz^2}{\mste^2} \KR \Bigg], \\
 && t = \frac{1}{2} \log \KL \frac{\mste^2 \mstz^2}{\mt^4} \KR .
\EEA

The second step of refinement incorporated into \fh\ concerns leading
QCD corrections beyond \twol\ order. They are taken into account by
using the $\overline{MS}$ top mass
\BE
\label{mtmsbar}
 \mtms = \mtms(\mt) \approx \frac{\mt}{1 + \frac{4}{3\,\pi}\als(\mt)} 
\EE
for the \twol\ contributions instead of the pole mass, $\mt = 175 \gev$.

\smallskip
The results implemented in \fh\ have been compared to
the calculations using RG methods. Good agreement has
been found in the case of no mixing in the $\Stop$ sector, i.e. 
$\Mtlr = 0 \gev$, whereas sizable deviations can occur when mixing in
the $\Stop$-sector is taken into account.
This has been discussed in detail in~\citere{mhiggs2lb}.


\subsection{Calculation of the $\rho$-parameter}

We have also implemented the calculation of the MSSM contributions to
$\De\rho$~\cite{drhosuqcd,precobssusyqcd}. Here the corrections 
arising from $\Stop/\Sbot$-loops up to $\oaas$
have been taken into account. The result is valid for arbitrary
parameters in  
the $\Stop$- and $\Sbot$-sector, also taking into account the mixing
in the $\Sbot$-sector which can have a non-negligible
effect in the large $\Tb$ scenario~\cite{precobssusyqcd}.  

The \twol\ result is separated into the pure gluon-exchange contribution,
which can be expressed by a very compact formula that allows a very
fast evaluation, and the pure
gluino-exchange contribution, which is given by a rather lengthy 
expression. The latter correction goes to zero with increasing gluino mass
and can thus be discarded for a heavy gluino%
\footnote{
An additional contribution to $\dr$, arising from a shift in the
squark masses when the soft SUSY breaking parameters are used as
input (due to the $SU(2)$ invariance of these parameters
in the squark sector), is not implemented in \fh.
This correction is described in detail
in \citere{drhosuqcd}.
}.
The $\rho$-parameter can be used as an additional constraint (besides
the experimental bounds) on the squark masses. 
A value of $\De\rho$ outside the experimentally preferred region of 
$\dr^{\SU} \approx 10^{-3}$~\cite{delrhoexp} indicates experimentally
disfavored $\Stop$- and $\Sbot$-masses.


\section{The Fortran program \fh}

\subsection{The main structure}

The complete program \fh\ consists of about 50.000 lines Fortran code,
where the main part belongs to the formulas for the renormalized Higgs
self-energies at the \twol\ level and the gluino contribution to
$\De\rho$. The executable file fills about 4 MB disk space.
The calculation for one set of parameters, including the $\De\rho$
constraint, with the highest accuracy takes less than 2 seconds on
a third-generation Alpha 21164 microprocessor (300 MHz processing
speed).

\smallskip
There exists a Home page for \fh:\\
{\tt http://www-itp.physik.uni-karlsruhe.de/feynhiggs}~.\\
Here a uu-encoded version as well as an ASCII version is available,
together with a short instruction, information about bug fixes etc.

\bigskip
\fh\ consists of several subprograms which are listed in
Table~\ref{fhproglist}. 
We now describe the different subprograms in detail.

\begin{table}[ht!]
\renewcommand{\arraystretch}{1.5}
\begin{center}
\begin{tabular}{|l||l|} \hline
{\bf subprogram} & {\bf function of the program} \\ \hline \hline
FeynHiggs.f      & front-end \\ \hline
FeynHiggsSub.f   & main part of \fh\ : calculation of $\mh,\mH$ \\ \hline
Hhmasssr2.f       & \onel\ self-energies \\ \hline
varcom.h         & definition of variables \\
bc.f             & \onel\ functions \\
lamspen.f        & further mathematical functions \\
def2.f           & definitions for the MSSM parameters \\ \hline
P1secode.f       & Fortran code for $\hSie_{\Pe}(0)$ \\
P1sesum.f        & putting together the different parts of
                   $\hSie_{\Pe}(0)$ \\
P1sevar.f        & definition of variables for $\hSie_{\Pe}(0)$ \\ \hline
P2se[code,sum,var].f      & same for $\hSie_{\Pz}(0)$ \\ \hline
P1P2se[code,sum,var].f    & same for $\hSie_{\PePz}(0)$ \\ \hline
P1setl[code,sum,var].f    & same for $\hSiz_{\Pe}(0)$ \\ \hline
P2setl[code,sum,var].f    & same for $\hSiz_{\Pz}(0)$ \\ \hline
P1P2setl[code,sum,var].f  & same for $\hSiz_{\PePz}(0)$ \\ \hline
delrhosub.f      & main part for the calculation of $\De\rho$ \\
delrhoGluino[code,sum,var].f & subroutines for the gluino-exchange
                               contribution to $\De\rho$ \\ \hline
\end{tabular}
\renewcommand{\arraystretch}{1}
\caption[]{The subprograms of \fh.}
\label{fhproglist}
\end{center}
\end{table}

\begin{itemize}

\item
{\bf FeynHiggs.f} 
is the front-end for the whole program. Here all
variables are set, all options are chosen, the subprogram 
{\bf FeynHiggsSub.f} is called, and the results for the Higgs masses are
printed out.
For an easy use of \fh\ this front-end can be manipulated at will,
the rest of the program need not to be changed.

\item
{\bf FeynHiggsSub.f}:
Here the actual calculation is carried out. The various results
for the 
renormalized Higgs-boson self-energies~(\ref{renhiggsse}) are put together.
Eq.~(\ref{higgsmasseq}) is solved numerically, the refinement terms (the
leading \twol\ Yukawa contribution and the leading QCD corrections
beyond \twol\ order) are incorporated.

\item
{\bf Hhmasssr2.f}:
The results for the complete \onel\ self-energies are evaluated.

\item
{\bf varcom.h}:
The variables needed for the files {\bf *code.f} are defined and
grouped in common blocks.

\item
{\bf bc.f, lamspen.f} 
contain mathematical functions needed for the
one- and \twol\ self-energies.

\item
{\bf def2.f}: 
The masses of the stops and sbottoms are calculated from the
parameters in the squark mass matrices. The opposite way is also
possible. 

\item
{\bf P1setlcode.f}
contains the complete code for the \twol\ contribution to the
renormalized $\Pe$ self-energy in 
$\oaas$. This code was first calculated with the help of \mma\ packages
(see~\cite{mhiggs2la,mhiggs2lb}) and was afterwards transformed
automatically into this Fortran code.
The complete code has been split up into 13 smaller functions in order not
to exceed the maximal number of continuation 
lines of the Fortran compiler.

\item
{\bf P1setlsum.f}:
The sum of 13 functions for the renormalized $\Pe$
self-energy at the \twol\ level, contained in {\bf P1setlcode.f}, is
put together. 

\item
{\bf P1setlvar.f}
contains the variable definition for the 13 subfunctions described
above. 

\item
The above three files exist in an analogous way for the other
renormalized Higgs-boson self-energies at the one- and at the \twol\
level. The \onel\ self-energies in these files are given in the
approximation explained in detail in~\cite{mhiggs2la,mhiggs2lb} and are
only needed internally for consistency checks. For the real
calculation of $\mh$ and $\mH$ the complete \onel\ self-energies are
calculated in {\bf Hhmasssr2.f}.

\item
{\bf delrhosub.f}
is needed for the calculation of the MSSM contribution to
$\De\rho$. In this program the evaluation of the leading \onel\
contribution is
performed. In addition also the gluon-exchange correction in $\oaas$
is calculated. The $\oaas$ contribution to $\dr$ is completed by
including the gluino-exchange contribution by calling the subprogram
{\bf delrhoGluinosum.f}.

\item
{\bf delrhoGluino[code,sum,var].f}:
The code for the gluino exchange contribution to $\De\rho$ is
implemented in the same way as it is described for the renormalized
Higgs-boson self-energies. 

\end{itemize}


\subsection{Options and setting of variables}

\fh\ can be run in several ways, determined by the choice of several
options. 

\begin{itemize}

\item
{\bf Depth of calculation} 
allows to choose to what extent the
refinements described in section~\ref{subsec:mhcalc} should be applied.

\item
{\bf Selection of input parameters:}
One can either use the physical parameters of the $\Stop$ mass matrix
($\mste, \mstz$ and $\tst$) or the soft SUSY breaking parameters 
($\MstL, \MstR$ and $\Mtlr$) as input parameters.

\item
{\bf $\mt$ in the $\Stop$ mass matrix:} 
If the soft SUSY breaking parameters are used as input parameters, one can
choose whether the top pole mass, $\mt$, or the running top mass,
$\mtms$, should be used in the $\Stop$ mass matrix in order to
determine the masses of the eigenstates $\mste, \mstz$ and
$\tst$. 

\item
{\bf Limit for $\De\rho^{\SU}$:}
One can specify the maximally allowed value for the MSSM contribution
to $\De\rho$. If $\De\rho^{\SU}$ exceeds this limit a warning is
printed out. 

\item
{\bf Selection for the \onel\ accuracy:}
Before the calculation of the Higgs masses starts, one has to
specify to what accuracy the \onel\ renormalized self-energies should
be evaluated. One can either take into account the top sector only,
one can choose to use the top {\em and} the bottom sector, or one can
select the option that the complete MSSM should be taken into account.

\end{itemize}

\begin{table}[ht!]
\renewcommand{\arraystretch}{1.5}
\begin{center}
\begin{tabular}{|c||c||c|} \hline
{\bf input for \fh} & {\bf expression in the MSSM} & 
{\bf internal expr. in \fh}
\\ \hline \hline
{\tt tan(beta)}          & $\Tb$      & {\tt ttb} \\
{\tt Msusy\_top\_L  }    & $\MstL$    & {\tt msusytl} \\
{\tt Msusy\_top\_R}      & $\MstR$    & {\tt msusytr} \\
                         & $\MsbL$    & {\tt msusybl} \\
                         & $\MsbR$    & {\tt msusybr} \\
{\tt MtLR }              & $\Mtlr$    & {\tt mtlr} \\
{\tt MSt2}               & $\mstz$    & {\tt mst2} \\
{\tt delmst}             & $\delmst = \mstz - \mste$ & {\tt delmst} \\
{\tt sin(theta\_stop)}   & $\sintt$   & {\tt stt} \\
                         & $\sintb$   & {\tt stb} \\
{\tt Mtop}~~or~~{\tt mt} & $\mt$      & {\tt mmt} \\
                         & $\mb$      & {\tt mbb} \\
{\tt Mgluino}            & $\mgl$     & {\tt mgl} \\
{\tt Mue}                & $\mu$      & {\tt mmue} \\
{\tt M}                  & $M$        & {\tt mmm} \\
{\tt MA}                 & $\MA$      & {\tt mma} \\
\hline
\end{tabular}
\renewcommand{\arraystretch}{1}
\caption[]{The meaning of the different MSSM variables which have to
be entered into \fh.}
\label{mssmparameters}
\end{center}
\end{table}

In {\bf FeynHiggs.f} the Standard Model (SM) variables are set to
their present experimental values. New values for these variables can
be implemented by the user into the code of the file {\bf FeynHiggs.f}
easily. 
The MSSM variables can be chosen by
the user at will. In addition, since the dependence of the Higgs
masses on the top mass is very strong, also $\mt$ can be chosen at
will. In Table~\ref{mssmparameters} the meaning of the different
parameters \fh\ asks for is explained%
\footnote{
Some MSSM variables exist internally in \fh, but are not input
parameters. These variables have no entry in the left column of
Table~\ref{mssmparameters}.
}.
All these variables are transferred to the different subprograms by
common blocks.


\section{How to use \fh}

In this section we will give two examples of how \fh\ is used.
As stated before, the front-end {\bf FeynHiggs.f} can be
manipulated by the user at will, whereas the subprogram {\bf
FeynHiggsSub.f} and all the other subprograms should not be
changed. 
Concerning a modificaton of the front-end one has to note that
all variables have to be defined in the front-end. 
The Higgs masses are then obtained by calling the subroutine via
\BEA
{\tt call~feynhiggssub(mh1,mh2,mh12,mh22)}~,\non
\EEA
where {\tt mh1} and {\tt mh2} are the \onel\ corrected values for
$\mh$ and $\mH$, respectively. {\tt mh12} and {\tt mh22} contain the
values for the \twol\ corrected Higgs masses, including the refinement
terms as it has been specified in the options.

\bigskip
In the following two examples are given, how the application of \fh\ looks
like, using the given front-end of the currently distributed
version. The user's input is given in  
{\tt {\bf bold face}} letters.

\newpage
\subsection{Example 1}

{\tt~>{\bf FeynHiggs.exe}\\
\dots~Introduction~\dots \\
~---------------------------------------------------\\
\\
~depth~of~calculation~?\\
~1:~full~1-loop~+~2-loop~QCD\\
~2:~same~as~1,~but~in~addition~with\\
\mbox{}~~~~~~~~~~~~~~mt~=~166.5~at~2-loop\\
~3:~same~as~2,~but~in~addition~with\\
\mbox{}~~~~~~~~~~~~~~Yukawa~term~added~for~light~Higgs\\
{\bf~1}\\
\\
~Select~input:\\
~1:~Msusy,~MtLR,~...\\
~2:~MSt2,~delmst,~stt,~...\\
{\bf~2}\\
\\
~Limit~for~Delta~rho~=~1.3~*~10\^~-3~?~(0~=~ok)\\
{\bf~0}\\
\\
\\
\\
~tan(beta)~=~?\\
{\bf~1.6}\\
\\
~MSt2~=~?\\
{\bf~400}\\
~delmst~=~?\\
{\bf~100}\\
~sin(theta\_stop)~=~?\\
~(0:~stt~=~0~//~1:~stt~=~sin(-Pi/4))\\
{\bf~1}\\
~Mtop~=~175~?~(0~=~ok)\\
{\bf~173.8}\\
~Mgluino~=~500~?~(0~=~ok)\\
{\bf~300}\\
~Mue~=~-200~?~(0~=~ok)\\
{\bf~100}\\
~M~=~?~(0:~M~=~400~//~1:~M~=~Msusy)\\
{\bf~1}\\

\noindent%
MA~=~?\\
{\bf~500}\\
\\
~Selection:~1~=~top~only,~2~=~top/bottom~only,~3~=~all\\
{\bf~3}\\
\\
~Your~parameters:\\
~tb,~Msusy(top-left),~Msusy(top-right),~MtLR\\
~MT,~Mgl,~Mue,~M,~MA\\
~~~1.600000~~~~~~~309.2813~~~~~~~308.0858~~~~~~~200.0000~\\
~~~173.8000~~~~~~~300.0000~~~~~~~100.0000~~~~~~~309.2813~~~~~~~500.0000\\
\\
~-------------------------------------------------\\
~The~results:~~light~Higgs~~~~~~heavy~Higgs\\
\\
~mh-tree~:~~~~~~~39.42879~~~~~~~~506.7153~~\\
~mh-1loop:~~~~~~~69.53253~~~~~~~~508.7314~~~\\
~mh-2loop:~~~~~~~64.10773~~~~~~~~508.3541~~~\\
~-------------------------------------------------\\
~Delta~rho~1-loop~~~~~~~~~:~~~5.909909368803866E-004\\
~Delta~rho~2-loop~(gluon)~:~~~6.177715322384914E-005\\
~Delta~rho~2-loop~(gluino):~~~4.755834893556049E-006\\
~Delta~rho~total~~~~~~~~~~:~~~6.575239249977918E-004\\
~-------------------------------------------------\\
\\
\\
~tan(beta)~=~?\\
\dots
}

In this example the physical parameters in the $\Stop$-sector have been
chosen as 
input parameters, no refinement term has been included. The selection
for $M$ sets $M = \MstL$, where $\MstL$ is calculated from 
$\mste$, $\mstz$ and $\sintt$, see \refeq{stopmassmatrix}.

\newpage
\subsection{Example~2}

{\tt~>{\bf FeynHiggs.exe}\\
\dots~Introduction~\dots~\\
~---------------------------------------------------\\
\\
~depth~of~calculation~?\\
~1:~full~1-loop~+~2-loop~QCD\\
~2:~same~as~1,~but~in~addition~with\\
\mbox{}~~~~~~~~~~~~~~mt~=~166.5~at~2-loop\\
~3:~same~as~2,~but~in~addition~with\\
\mbox{}~~~~~~~~~~~~~~Yukawa~term~added~for~light~Higgs\\
{\bf~3}\\
\\
~mt~in~the~stop~mass~matrix~at~2-loop~?\\
~1:~mt~=~pole~mass\\
~2:~mt~=~running~mass\\
{\bf~2}\\
\\
~Select~input:\\
~1:~Msusy,~MtLR,~...\\
~2:~MSt2,~delmst,~stt,~...\\
{\bf~1}\\
\\
~Limit~for~Delta~rho~=~1.3~*~10\^~-3~?~(0~=~ok)\\
{\bf~0.00001}\\
\\
\\
\\
~tan(beta)~=~?\\
{\bf~20}\\
\\
~Msusy\_top\_L~=~?\\
{\bf~1000}\\
~Msusy\_top\_R~=~?~(0:~Msusy\_top\_R~=~Msusy\_top\_L)\\
{\bf~0}\\
~Mtop~=~175~?~(0~=~ok)\\
{\bf~0}\\
~Mgluino~=~500~?~(0~=~ok)\\
{\bf~0}\\
~Mue~=~-200~?~(0~=~ok)\\
{\bf~-100}\\

\noindent%
M~=~?~(0:~M~=~400~//~1:~M~=~Msusy)\\
{\bf~300}\\
~MA~=~?\\
{\bf~400}\\
~MtLR~=~?\\
{\bf~1000}\\
\\
~Selection:~1~=~top~only,~2~=~top/bottom~only,~3~=~all\\
{\bf~3}\\
\\
~Your~parameters:\\
~tb,~Msusy(top-left),~Msusy(top-right),~MtLR\\
~MT,~Mgl,~Mue,~M,~MA\\
~~~20.00000~~~~~~~1000.000~~~~~~~1000.000~~~~~~~1000.000~\\
~~~175.0000~~~~~~~500.0000~~~~~~-100.0000~~~~~~~300.0000~~~~~~~400.0000\\
\\
~-------------------------------------------------\\
~The~results:~~light~Higgs~~~~~heavy~Higgs\\
\\
~mh-tree~:~~~~~~90.70748~~~~~~~400.1090~~\\~~
~mh-1loop:~~~~~~133.6753~~~~~~~400.1799~~\\~~
~mh-2loop:~~~~~~118.8657~~~~~~~400.1770~~\\~~
~-------------------------------------------------\\
~using~running~mt~for~two-loop~contribution:~~~167.338636349986\\
~...~also~for~mt~in~Stop~mass~matrix\\
~mh-2loop:~~~~~~120.8774~~~~~~~400.1780\\
~-------------------------------------------------\\
~using~running~mt~for~two-loop~contribution:~~~167.338636349986~\\
~...~also~for~mt~in~Stop~mass~matrix\\
~adding~Yukawa~term~for~light~Higgs\\
~...~also~with~running~mt~in~stop~mass~matrix\\
~mh-2loop:~~~~~~122.2504~~~~~~~400.1780~~\\
~-------------------------------------------------\\
~WARNING:~Delta~rho~>~experimental~limit\\
~Delta~rho~1-loop~~~~~~~~~:~~~3.224156596235517E-005\\
~Delta~rho~2-loop~(gluon)~:~~~3.475299800114144E-006\\
~Delta~rho~2-loop~(gluino):~~~2.896993903992095E-005\\
~Delta~rho~total~~~~~~~~~~:~~~6.468680480239026E-005\\
~-------------------------------------------------\\
\\
\\
~tan(beta)~=~?\\
\dots
}

\newpage
If one selects to include all refinement terms, the front-end
automatically calculates the Higgs masses in all three steps of
accuracy. The running top mass (in this example
{\tt 167.338636349986}~GeV) is used
also in the $\Stop$ mass matrix. 
In this example only a very small SUSY contribution to $\dr$ ia
allowed and the value of $\De\rho^{\SU}$ for the chosen parameters
exceeds the above specified maximal value (a warning is printed out.)


\section{Conclusions}

\fh\ is a Fortran code for the calculation of the masses of the
neutral $\cp$-even Higgs bosons of the MSSM. 
It is based on results which have been obtained using
the Feynman-diagrammatic approach. The results consist of the full
\onel\ and the leading \twol\ QCD corrections.
Two further steps of refinements have been implemented. 
The Fortran code for the \onel\ contribution has been taken over from a
program written by A.~Dabelstein. 
The Fortran code for the \twol\ correction has 
been generated automatically from a \mma\ result.

The program is available via the WWW page\\
{\tt http://www-itp.physik.uni-karlsruhe.de/feynhiggs}~.

The code consists of a front-end and a subroutine. The front-end can
be manipulated at the user's will. The different parts
of the code have been described in detail, and the meaning of the
variables used 
in the code has been explained. We have given two examples of how to 
use \fh.

The subroutine is self-contained and can be linked as an independent
part to other existing programs%
\footnote{
This has already successfully been performed for a Fortran program
used by members of the DELPHI collaboration at
Karlsruhe~\cite{sopczak}.
}
thus providing a precise prediction for $\mh$ and $\mH$, which can
then be used for further computations.

\bigskip
\subsection*{Acknowledgements}
We thank A.~Dabelstein for providing his \onel\ program for the
calculation of the Higgs-boson masses.
W.H. gratefully acknowledges support by the Volkswagenstiftung.




\end{document}